\documentclass[useAMS,usegraphicx,usenatbib]{mn2e} 
\usepackage{amsmath,graphicx}
\usepackage{subfigure}
\usepackage{ulem}
\usepackage{color}
\usepackage{graphicx}
\usepackage{times} 
\usepackage{amssymb}
\usepackage{amsmath}
\usepackage{lscape}
\usepackage{url}
\newif\ifAMStwofonts
\AMStwofontstrue

\def\rg{${\it r}_{\rm g}$}
\def\rin{${\it r}_{\rm in}$}
\def\laor{\rm{\sc LAOR}}

\def\diskbb{\rm{\sc DISKBB}}
\def\diskpn{\rm{\sc DISKPN}}

\def\nh{${\it N}_{\rm H}$}
\def\ka{$K\alpha$}

\def\epicmos1{{\it EPIC}{\rm-MOS1}}
\def\epicmos2{{\it EPIC}{\rm-MOS2}}
\def\epicmos{{\it EPIC}{\rm-MOS}}

\def\chandra{{\it Chandra}}

\def\asca{{\it ASCA}}

\def\rxte{{\it RXTE}}

\def\xspec{\hbox{\sc XSPEC}}
\def\xspecv{{\sc XSPEC}{\rm\thinspace v\thinspace 12.4.0}}
\def\ciao{\hbox{\rm CIAO}}

\def\ks{\hbox{$\rm\thinspace ks$}}

\def\deg{$^{\circ}$}  

\def\cm{\hbox{$\rm\thinspace cm$}}

\def\kpc{\hbox{$\rm\thinspace kpc$}}

\def\ev{\hbox{$\rm\thinspace eV$}}
\def\kev{\hbox{$\rm\thinspace keV$}}

\def\msun{\hbox{$\rm\thinspace M_{\odot}$}}

\def\gx{\hbox{\rm GX 339-4}}

\def\j{\hbox{\rm XTE J1118+480}}

\begin{document}

\title[Soft component in the LHS of \j] {Thermal emission from the stellar-mass black hole binary \j\
  in the low/hard state. } \author[Reis, Miller, \& Fabian]
{\parbox[]{6.in} {R.~C.~Reis $^{1}$\thanks{E-mail:
      rcr36@ast.cam.ac.uk}, J.~M. Miller$^2$ and A.~C.
    Fabian$^{1}$\\ } \\
  \footnotesize
  $^{1}$Institute of Astronomy, Madingley Road, Cambridge, CB3 0HA\\
  $^{2}$Department of Astronomy, University of Michigan, 500 Church
  Street, Ann Arbor, MI 48109, USA}

\maketitle

\begin{abstract}
  We report on the detection of a thermal-disk component from the
  stellar-mass black hole binary \j\ in the canonical low/hard
  state. The presence of a thermal component with a temperature of
  approximately 0.21\kev\ in the \chandra\ spectra of \j\ is found at
  more than the 14$\sigma$ confidence level. Based on this evidence we
  argue that the accretion disk in \j\ is not truncated far from the
  central black hole in contrast with previous claims.

\end{abstract}


\section{Introduction}

The X-ray spectra of Galactic black hole binaries conveys important
information on the geometry of the accretion disk surrounding the
black hole. These sources have usually been characterised by the
relative strength of their soft and hard X-ray emission. In the
thermal or High/soft state (HSS; see McClintock \& Remillard 2006) the
soft spectrum is dominated by thermal emission thought to originate in
a standard thin accretion disk extending to the innermost stable
circular orbit (ISCO; Shakura \& Sunyaev 1973).  The presence of a
powerlaw continuum is faint and quasi-periodic oscillations (QPOs) are
absent or very weak. In the steep powerlaw or very high state the
thermal disk still dominates the soft X-ray emission but now the power
law component is much more distinguishable and QPOs are observed
(Remillard \& McClintock 2006). A reflection component
(Ross \& Fabian 1993; Miller 2007) predominantly around the Fe-\ka\
line emission (6.4--6.97\kev) has also been observed in various
systems in these states (Miniutti et al. 2004; Miller et al. 2004,
2008; Reis et al. 2008, 2009).  Reflection features in the spectra of
these HSS and VHS system arise as hard emission, possibly from a
corona, irradiates the cooler, optically thick disk below and give
rises to fluorescent and recombination emission.

The geometry of the accretion flow in the low/hard state (LHS) however
remains a topic of debate. In the accretion disk corona model for the
LHS the hard X-ray emission originates either in the base of a jet
(Merloni \& Fabian 2002; Markoff \& Nowak 2004; Markoff, Nowak \&
Wilms 2005) or due to magnetic flares in an accretion disk corona
where the X-rays are produced by inverse Compton scattering of soft
photons (Merloni, Di Matteo \& Fabian 2000; Merloni \& Fabian
2001). In both these cases the corona surrounds a thin accretion disk
possibly extending close to the ISCO.  An alternative model has the
accretion flow consisting of a thin disk truncated at large distances
from the black hole (Esin et al. 1997, 2001). In this scenario the
central region is filled with a hot, quasi-spherical
advection-dominated accretion flow (ADAF) and the accretion disk
temperature should peak in the far ultraviolet where interstellar
absorption usually complicate its direct detection.

Due to its unusually high Galactic latitude ($b=+62$\deg) and low
interstellar  absorption, \j\  has been  the focus  of multiwavelength
studies  (Hynes  et   al.  2000;  Frontera  et  al.   2001;  Chaty  et
al. 2003).  Dynamical measurements of  \j\ set a strong  constraint on
the  mass-function   of  $6.1\pm  0.3$\msun\  (Wagner   et  al.  2001;
McClintock et al.  2001). Recent studies suggests that  the black hole
in \j\ has a mass of $8.53\pm0.60$\msun\ and an orbital inclination of
$68\pm2$\deg\ (Gelino et al. 2006).  The same authors place the system
at a distance  of $1.72\pm 0.10$\kpc\ in agreement  with that previously
suggested  by   McClintock  et   al.  (2001).  However,   much  higher
inclinations  have  been reported  by  other  studies  with Wagner  et
al. (2001)  reporting a  value of 81\deg$\pm2$\deg,  Zurita et  al. (2002)
constraining  it to  71--82\deg\  and more  recently  Khruzina et  al.
(2005) with $i=80^{+1}_{-4}$ degrees.

XTE{\thinspace J1118+480} was observed in its LHS by \chandra\ in 2000
as part of a multiwavelength, multiepoch observing campaign. Based on
these observation, McClintock et al. (2001) reported an apparent cool
thermal component at $\approx 24$\ev\ which was interpreted as being
caused by a truncated accretion disk with $R_{{\rm tr}}\gtrsim70$\rg\
(\rg$=GM/c^2$), much larger than the expected ISCO. This motivated an
ADAF interpretation for the system in \j\ which was presented in a
later paper by Esin et al. (2001). The cool thermal component reported
by McClintock et al. (2001) was in contrast with a previous \asca\
observation where in an IAU telegram, Yamaoka et al. (2000) detected a
blackbody component with a temperature of $\approx 0.2$\kev. This
temperature is characteristic of a disk approaching the ISCO in the
low/hard state of X-ray binaries (see e.g. Miller et al. 2006; Miller
2007; Miller et al. 2008; Reis et al. 2008, 2009). In this paper we
show that a soft-thermal disk component with a temperature similar to
that reported by Yamaoka et al. (2000) is clearly present in the
\chandra\ 2000 observation and is consistent with a disk extending
close to the innermost stable circular orbit. In the following
sections we detail our analysis procedure and results.

\section{Observation and Data reduction}

We analysed the 2000 April 18 \chandra\ and \rxte\ observation of \j\
in its low/hard state. \j\ was observed with the {\it Low Energy
  Transmission Grating Spectrometer} (LETGS, Brinkman et al. 2000) and
the ACIS-S detector on board \chandra\ for an integrated exposure of
27.2\ks\ and simultaneously by \rxte\ for a total combined exposure of
3.5\ks\ (OBS ID 50407-01-02-03 and 50407-01-02-04).{\footnote {In this
    work we are only using the two \rxte\ observation that directly
    overlapped with that of \chandra. For a detailed analysis using
    all of the \rxte\ observation made between 2000 13 April and 15
    May see Miller et al. (2002).}} The positive and negative
first-order \chandra\ LETG spectra were extracted following the
Science Threads for Grating Spectroscopy in the \ciao\ 4.0 data
analysis software{\footnote{Chandra Interactive Analysis of
    Observation (\ciao), Fruscione et al. 2006;
    http://cxc.harvard.edu/ciao/threads/gspec.html}}. The nominal
LETGS/ACIS-S energy coverage is between 0.20--10\kev\ (Weisskopf
2004), however the spectrum is noisier above 7\kev\ and below
$\approx0.4$\kev. For this reason we follow the restriction imposed by
McClintock et al. (2001) and Miller et al. (2002) and restrict
spectral analysis of the LETGS data to the energy range 0.3--7.0\kev.

\rxte\ data were reduced in the standard way using the {\rm HEASOFT
  v6.6.0 } software package. We used the ``Standard 2 mode'' data from
the {\it Proportional Counter Array} (PCA) using PCUs 2 and 3 as well
as 64 channel data collected with the {\it High energy X-ray Timing
  Experiment} (HEXTE) in cluster A and B. For the PCA and HXTE spectra
we restrict our analyses to the 2.8--25 and 20--100\kev\ energy range
respectively. To account for residual uncertainties in the calibration
of PCU 0 and 2, we added a 0.6 per cent systematic error to all energy
channel.  In the initial analyses of the \chandra\ data McClintock et
al. (2001) used a custom IDL software and a preliminary spectral
response, therefore we cannot reproduce their results. Furthermore the
authors do not require a minimum number of counts per energy bin. This
differs in the present work where we use the latest response software
and the {\it FTOOL} {\rm grppha} to require at least 20 counts per
energy bin so as to enable the use of $\chi^2$ statistics in all our
analyses. All parameters in fits involving different instruments were
tied and a normalisation constant was introduced. \xspecv\ (Arnaud
1996) was used to analyse all spectra. The quoted errors on all
derived model parameters correspond to a 90 per cent confidence level
for one parameter of interest ($\Delta\chi^2=2.71$ criterion) unless
stated otherwise.

\begin{figure}
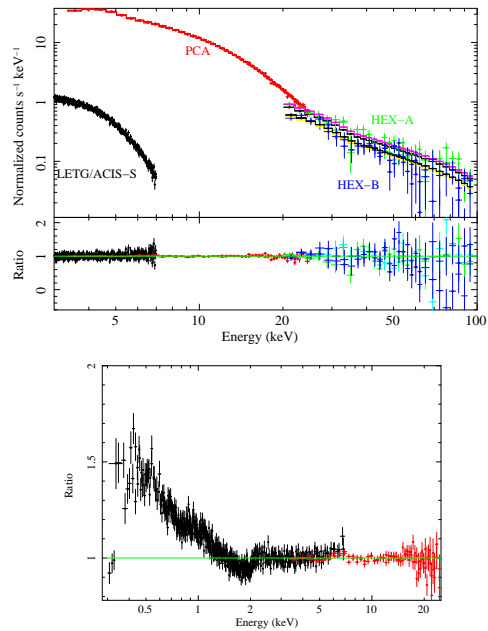

\begin{center}
\rotatebox{270}{
\resizebox{!}{6.5cm} 
{\includegraphics{figure_po_above3kev.ps}}
}
\rotatebox{270}{
\resizebox{!}{5.cm} 
{\includegraphics{ratio_thermal.ps}}
}
\end{center}
\caption{{\it Top} \rxte\ and \chandra\ spectra of \j\ fitted
  simultaneously above 3\kev. The data were fitted with a single
  powerlaw component modified by absorption in the interstellar medium
  (\nh=$1.3\times10^{20}{\rm cm^{-2}}$). {\it Bottom}: Data/model plot
  for the full energy range 0.3--100\kev\ showing the presence of a disk
  component. {\rm HEXTE} data omitted for
  illustration purposes only }

\end{figure}

\section{analysis and results}

Anticipating a power-law modified by interstellar absorption ({\rm
  PHABS}{\footnote {Using the standard BCMC cross-sections
    (Balucinska-Church and McCammon 1992) and ANGR abundances (Anders
    \& Grevesse 1989).} model in \xspec) to provide a good fit to most
  of the energy range under consideration, we began by looking at the
  \rxte\ data in conjunction with \chandra\ data above 3\kev. This low
  energy cutoff was chosen as we do not expect major contribution from
  any black-body component at these energies for a source in the
  LHS. With the value of the power-law index tied between the
  \chandra\ and \rxte\ observations, and an equivalent neutral column
  density fixed at $1.3\times10^{20}{\rm cm^{-2}}$ as suggested by
  McClintock et al. (2001), we obtain an excellent fit with
  $\chi^2/\nu= 578.6/583$ and a photo-index of $\Gamma= 1.756 \pm
  0.006$ in agreement with that reported by McClintock et al. (2001).
  Figure 1 (Top) shows the data fitted with a powerlaw above 3\kev. By
  extending the \chandra\ energy range to that of 0.3-7.0\kev\ it is
  clear that this single powerlaw does not provide a good fit to the
  full energy range and a strong soft excess is seen (Fig. 1 bottom)
  in disagreement with the results presented by McClintock et
  al. (2001). A single absorbed-powerlaw yields an unsatisfactory fit
  to the full 0.3--100.0\kev\ energy range, with $\chi^2/\nu=
  6698.6/4461$.  Allowing the column density to vary between
  1.0--1.3$\times10^{20}{\rm cm^{-2}}$ as per McClintock et al. (2001)
  marginally improved the fit with $\chi^2/\nu= 6461.4/4460$.  Figure
  2 shows the residuals to this fit in the full energy range. Allowing
  the column density to vary over a wider range{\footnote{The lower
      and upper limits are derived from maps of IR emission (Schlegel
      et al. 1998) and Ca {\rm II} absorption features (Dubus et
      al. 2001) respectively.}}  (0.67--2.8$\times10^{20}{\rm
    cm^{-2}}$) yields a similar unsatisfactory result with
  $\chi^2/\nu=6240.7/4460$.

  To model this soft excess we initially used the multicolour disk
  blackbody model \diskbb\ (Mitsuda et al. 1984 ; Makishima et
  al. 1986). The neutral hydrogen column density was constrained to
  vary between 1.0--1.3$\times10^{20}{\rm cm^{-2}}$ in accord with the
  most likely range reported by McClintock et al. (2001). This
  resulted in a much improved overall fit with $\chi^2/\nu=
  4244.6/4458$ and an F-test value of 1164.14 (1048.23 when \nh\ is
  allowed to vary over a wider range) over the fit without
  \diskbb. With a disk normalisation of $5800\pm400$ (1$\sigma$
  confidence level) the presence of this thermal disk is thus
  effectively confirmed at more than the 14$\sigma$ level. The best
  fit resulted in an effective disk temperature of
  $0.203\pm0.005$\kev. This is in agreement with the value reported by
  Yamaoka et al. (2000) where a black-body component with a
  temperature of $0.2\pm0.1$\kev\ was found in an \asca\ observation
  of \j.  However, there appears to be an instrumental artifact at an
  energy of approximately 2\kev\ (see bottom panel of Fig. 1 and
  Fig. 2) consistent with an edge due to the Iridium coating of the
  detector. Following Miller et al. (2002), we modelled this using an
  inverse edge at an energy of $\approx2$\kev ($\tau <-0.1$). This
  further improved the quality of the fit with $\chi^2/ \nu=
  4133.8/4456$. The various parameters for this fit are shown in Table
  1.{\footnote {A possible feature around 6.4\kev\ in the \rxte\
      spectra can be seen in Fig. 2. Adding a Gaussian at 6.4\kev\
      corresponding to iron fluorescence emission improves the fit to
      $\chi^2/\nu= 4075.6/4454$. This broad line ($\sigma\sim1\kev$)
      has an equivalent width of approximately 100\ev. Replacing the
      Gaussian with a \laor\ line results in a further improvement to
      the fit with $\chi^2/\nu= 4043.9/4453$ and an inner radius
      characteristic of a maximally rotating black hole
      (\rin$\sim1.6$\rg).}  In order to investigate any degeneracy
    between the thermal components (disk normalisation and
    temperature) and the column density we explored their {\it full}
    parameter space using the ``contour'' command in \xspec\ with all
    parameters free to vary. Fig. 3 shows the 68, 90 and 99 per cent
    contour for two parameters of interest. With the column density
    allowed to vary over its full parameter space the best fit value
    approaches 2.7$\times10^{20}{\rm cm^{-2}}$, significantly higher
    than the range suggested by McClintock et al. (2001). However the
    disk temperature remains above $\sim0.18$\kev\ at the 99 per cent
    confidence level (Fig. 3).

 \begin{figure}
\centering
\rotatebox{270}{
\resizebox{!}{6cm} 
{\includegraphics{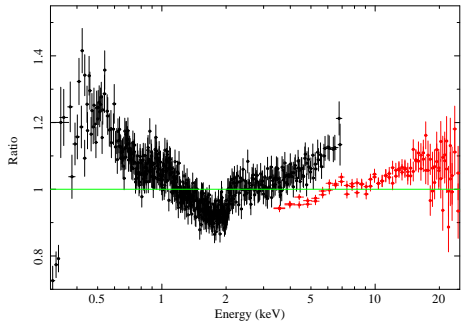}}
}
\caption{ Residuals to a fit with a simple powerlaw over the full
  0.3-100.0\kev\ energy range ({\rm HEXTE} data omitted for
  illustration purposes only). It is clear that a simple powerlaw does
  not provide a possible fit to the full energy range. }

\end{figure} 

\begin{figure}
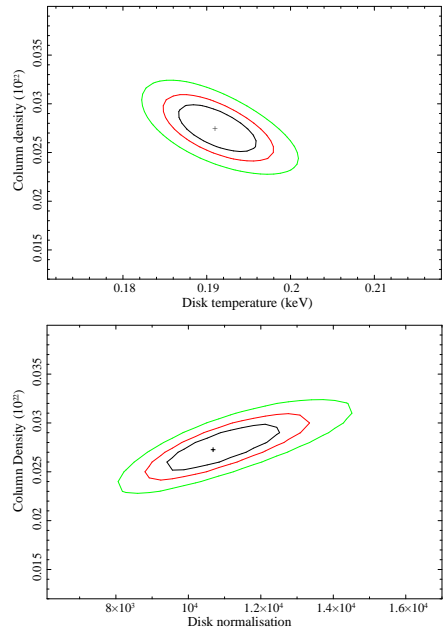

\centering
{
 \rotatebox{270}{
\resizebox{!}{5.7cm} 
{\includegraphics{contourky_nh.ps}  
}}}
\hspace{1cm}
{  
  \rotatebox{270}{
\resizebox{!}{5.7cm} 
{\includegraphics{contournorm_nh.ps}} 
}}

\caption{ Contour plot of hydrogen column density versus ({\it Top}) disk
  temperature and ({\it Bottom}) disk normalisation for \j. The 68, 90 and
  99 per cent confidence range for two parameters of interest are
  shown in black, red and green respectively. It can be seen that even
  when the column density is allowed to vary over its full parameter
  space the disk temperature remains greater than 0.18\kev\ at the 99
  per cent confidence level. }

\end{figure}

\begin{table}
\begin{center}
  \caption{ Results of simultaneously fitting the \chandra\ and
    \rxte\ data with a powerlaw and thermal disk component. It can be
    seen that statistically we cannot differentiate a model with an
    inner radius of 6 or 70\rg. }

\begin{tabular}{lcccccccccc}                
\hline
\hline
Model  & \diskbb   &\diskpn &   \diskpn\\
\hline
\nh  & $1.30_{-0.02}$              & $1.30_{-0.02}$           &  $1.30_{-0.02}$ \\ 
\rin( \rg)                & --                      & 6(f)                 &   70(f)  \\
kT  (\kev)                &$0.213\pm0.005$    & $0.203\pm0.005$  & $0.216\pm0.005$   \\
$N_{MCD}$                 & $4900\pm500$       &$0.088^{+0.010}_{-0.009}$    &$0.0038\pm0.0004$  \\
$\Gamma$                 &$1.729\pm0.005$            & $1.729\pm0.005$   &$1.729\pm0.005$   \\ 
$N_{pl}$                  &$0.1919\pm0.0015$  &$ 0.1919\pm0.0015$ & $ 0.1918\pm0.0015$ \\
$\chi^{2}/\nu$            &  4133.8/4456            &     4135.6/4456         &   4136.7/4456     \\

\hline
\hline
\end{tabular}
\end{center}

\small Notes.- All fits
contain an inverse edge with an energy of $2.09\pm0.01$\kev\ and $\tau
=-0.08\pm0.01$ as described in the text. The various models are described in \xspec\ as {\rm PHABS}$\times$({\rm PL + \diskpn}) or \diskbb. Error refers to the 90 per cent confidence range. The normalisation of each component is referred to as $N$. The column density \nh\ is in units of  $(10^{20} \cm^{-2})$ .

\end{table}

\begin{figure}
\begin{center}
\rotatebox{270}{
\resizebox{!}{6.5cm} 
{\includegraphics{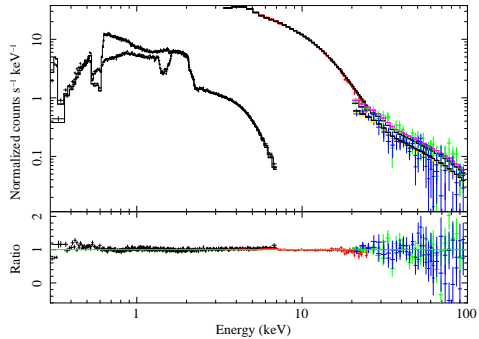}}
}
\end{center}
\caption{\rxte\ and \chandra\ spectra of \j\ fitted simultaneously for
  the full energy range. The data were fitted with a single powerlaw
  component modified by absorption in the interstellar medium and a
  multicolour disk blackbody with a temperature of $\approx
  0.21$\kev. The data have been rebinned for plotting purposes only. }

\end{figure}

Having shown above that the presence of a disk component is required
at more than the 14$\sigma$ level, we investigated the possibility of
constraining the innermost radius of emission. To do this we used the
\xspec\ model \diskpn\ (Gierlinski et al. 1999) which is a modified
version of \diskbb\ where the torque-free inner boundary condition is
taken into account. This model has three parameters: The maximum
colour temperature of the disk ($T_{col}$) in units of \kev, the inner
disk radius, \rin, and the normalisation which is defined as ${\rm
  m^2cos}i/d^2\beta^4$, where {\rm m} is the mass of the black hole in
solar masses, {\rm d} is the distance to the source in \kpc\ and
$\beta$ is the colour correction factor.  We performed a fit on the
full energy range with the inner radius, \rin\ fixed at both the value
expected for a disk extending down to the innermost stable circular
orbit of a non-spinning black hole (6\rg) and that of the truncated
disk predicted by McClintock et al. (2001) of 70\rg. Table 1
summarises our results and Fig. 4 shows the best fit spectra with the
multicolour disk black body having an inner radius fixed at 6\rg. 

It can be seen from Table 1 that models with and without a truncated
disk gives equally satisfactory fits. In order to differentiate
between these two models we have to consider the physical significance
of their respective parameters.  The mass and distance to the black
hole in \j\ has recently been estimated at $8.53\pm0.60\msun$ and
$1.72\pm0.1$\kpc\ respectively (Gelino et al. 2006). The inclination
of the system however remains highly uncertain with various studies
placing an upper limit of $\approx 83$\deg\ (Wagner et al. 2001;
Zurita et al. 2002; Khruzina et al. 2005). Using these values in
conjunction with the values obtained for the normalisation of the
\diskpn\ models in Table 1 we see in Fig. 5 that for a disk truncated
at 70\rg, the inclination lies below 83\deg\ only when the colour
correction factor $\beta\gtrsim5$. For the disk extending to 6\rg, the
colour correction factor, $\beta$ only needs to be greater than
$\approx2.2$ in order for the derived inclination to lie below 83
degrees. It was shown in Merloni, Fabian \& Ross (2000) that the
colour correction factor $\beta$, which need not be constant, varies
between $1.7<\beta<3$. With this limitation on $\beta$ it is highly
unlikely that the disk in \j\ is truncated at 70\rg\ and could even
extend within 6\rg\ indicating the presence of a rotating Kerr black
hole.

\section{Discussion}

A thermal component with a temperature of approximately $0.21$\kev\ in
the \chandra\ spectra of \j\ has been found at more than the
14$\sigma$ confidence level. This component has already been reported
in a previous observation of the source with \asca\ (Yamaoka et
al. 2000) but has nonetheless been overlooked in previous analyses of
the \chandra\ observation. Previous claims that the accretion disk in
\j\ is truncated at a radius greater than 70\rg\ is based on the lack
of evidence for this soft thermal component.

In the previous section we have shown that a disk truncated at both 6
and 70\rg\ gives equally satisfactory fits to the current
data. However, based on the current upper limit on the inclination of
the system ($\approx83$\deg), the model containing a disk truncated at
70\rg\ seems unphysical. Furthermore, with a disk temperature of
$\approx 0.21$\kev\ it is likely that the disk in \j\ is {\it not}
truncated far from the black hole.  This temperature is broadly
consistent with the $L_x\propto T^4 $ relation expected for a
geometrically stable blackbody. For \j\ at a distance of $\approx
1.8$\kpc\ and a mass of $\approx 8\msun$ we obtain a blackbody radius
extending to $<6$\rg. Similar results have been presented for the
black hole candidate {\rm XTE J1817--330} (Rykoff et al. 2007) where
the authors have followed the evolution of the system from the
high-soft state through to the low/hard state and found that the disk
did not recede after the state transition. Contrary to this
interpretation, Gierlinski et al. (2008 suggested that irradiation of
a truncated accretion disk by Comptonized photons could lead to a disk
radius underestimated by a factor of 2--3. It must be noted however
that the prescription of this model, specially in the case of {\rm XTE
  J1817--330}, requires that the irradiation somehow reproduces the
$L_x\propto T^4 $ relation expected for a simple black body observed
by Rykoff et al. (2007). Furthermore, in spectra where we observe both
a broad Fe-\ka\ line as well as the disk continuum, the shape of the
broad line argues strongly against a recessed disk, as is the case for
the black hole binary \gx\ in its low/hard state (Miller et al. 2006,
Reis et al. 2008). However, it must be noted that in both the present
work on \j\ and that of Rykoff et al. (2007) on {\rm XTE J1817--330},
$L_x/L_{Edd}\gtrsim 0.001$ which prompts the question of whether this
could be a bright phase of the low/hard state. It is still possible
that an advective flow take over at some point below $L_x/L_{Edd}\sim
0.001$.

\begin{figure}
\begin{center}
\rotatebox{270}{
\resizebox{!}{6.5cm} 
{\includegraphics{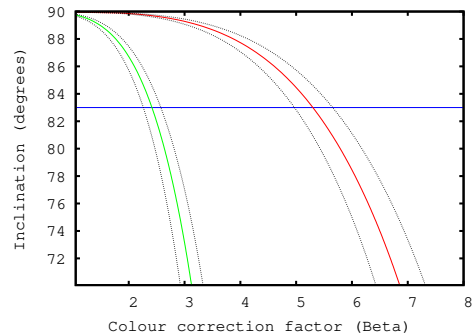}}
}
\end{center}
\caption{Inclination vs colour correction factor ($\beta$). The red
  and green curves are for a disk with inner radius of 70 and 6\rg\
  respectively. The dashed black lines show the error range. The upper
  limit for the inclination in \j\ is shown by the solid blue line.  }

\end{figure}

The temperature of this putative disk is very tightly constrained to
approximately $0.21$\kev\ and is independent of the MCD model used
(Table 1). Using the measured X-ray flux of the system (non-absorbed
flux in the 0.1--200\kev\ range) and estimates of its distance and
mass (McClintock et al. 2001; Gelino et al. 2006) we can make an
estimate on the upper limit on the radius of the accretion disk based
on a colour temperature of $0.21$\kev. The radial dependence of the
effective temperature of an accretion disk is found to be
$T^{eff}_{(R)}= T_{col}/\beta= (3GM \dot{M} f / 8\pi \sigma_T
R^{3})^{1/4} $ (Frank, King \& Raine 1992), where $f=
1-(R_{in}/R)^{1/2}$ and $\dot{M}= 4\pi D^2 F_x {\rm cos} i/ \epsilon
c^2$. With ${\rm F}=F_{x}/10^9{\rm ergs{\thinspace
    cm^{-2}}s^{-1}}\approx 5$, ${\rm r}=R/{\rm r_g}$, ${\rm
  m}=M/\msun$ and ${\rm d}=D/\kpc$, we obtain $T^{eff}_{({\rm r})}
\approx 1.34 ({\rm d}^2{\rm cos} i /\epsilon{\rm m}^2)^{1/4} (f/ {\rm
  r}^3)^{1/4}$. Assuming a Schwarzchild black hole with an efficiency
of 6 per cent, mass of 8.5\msun, and a distance of 1.72\kpc, as well
an upper limit on the inclination of 83 degrees and $\beta<3$ results
in an upper limit for the disk radius of $\approx16$\rg. Note that
this value is obtained in the extreme case of $\beta=3$. Lowering
$\beta$ to 2.4 results in a decrease in the upper limit of the disk
radius of $\approx10$\rg. We further note that the presence of a broad
fluorescence line in the \rxte\ data corroborates the existence of a
dense disc extending close to the black hole.

If the disk does indeed extend close to the ISCO in \j, it is
plausible that the broadband spectral energy distribution (SED)
observed in various multiwavelength campaign (Hynes et al. 2000;
Frontera et al. 2001; Chaty et al. 2003) is mostly due to
optically-thin synchrotron emission such as that originating in the
innermost part of a jet (Markoff et al. 2001) or magnetic flares in
the inner part of the accretion flow (Merloni, Di Matteo \& Fabian
2000). The strong UV hump observed in this system would thus not be
exclusively due to viscous dissipation from a cold, truncated
accretion disk. Another possibility would be that this optical-UV
emission is dominated by reprocessed hard X-ray emission (King \&
Ritter 1998) as is likely for {\rm XTE J1817--330} (Rykoff et
al. 2007). However, based on fast X-ray and optical variability,
Kanbach et al. (2001) argued against reprocessing and favoured
synchrotron emission as the source of the optical and UV emission in
\j. This view was further strengthened by Hynes et al. (2003; 2006)
who, on the basis of the rapid X-ray, UV, optical and infrared
variability, argue that the SED in \j\ is mostly dominated by
synchrotron emission possibly originating at the base of a jet. The
same authors also argued that a cold, thermal-disk component cannot
satisfactorily model the IR variability (Hynes et al. 2006). Noting
the presence of simultaneous quasi-periodic oscillation in both the
optical and X-ray light curves of \j, Merloni, Di Matteo \& Fabian
(2000) suggested that magnetic flares in an accretion disk corona
possibly extending close to the ISCO could explain the broadband SED
in \j. Our results further support the notion that the broadband SED
in \j\ is caused by inverse Compton scattering of soft photons in a
corona embedding a thin accretion disk extending close to the ISCO.

\section{Conclusions}

We have studied the \chandra\ observation of the stellar mass black
hole binary \j\ in the canonical low/hard state. A thermal disk
emission with a temperature of approximately $0.21$\kev\ is found at
greater than the 14$\sigma$ level. For \j\ this thermal emission most
likely originates from an accretion disk extending close to the radius
of marginal stability. The presence of a disk component in the
\chandra\ spectra of \j\ has been overlooked in previous analysis,
which resulted in \j\ becoming the archetype for ADAF scenarios. In
light of our analysis this picture needs to be reconsidered.

\section{Acknowledgements}
We thank Mike Nowak and Phil Uttley for useful advice and comments on
the draft paper. RCR acknowledges STFC for financial support. ACF
thanks the Royal Society.

\end{document}